\documentclass[letterpaper,english,aps, prb, reprint, superscriptaddress]{revtex4-1}
\usepackage[T1]{fontenc}
\usepackage[latin9]{inputenc}
\usepackage{amsmath}
\usepackage{amssymb}
\usepackage{bbold}
\usepackage{graphicx}

\makeatletter

\pdfpageheight\paperheight
\pdfpagewidth\paperwidth

\PassOptionsToPackage{position=top}{subfig}
\usepackage{hyperref}
\hypersetup{
	colorlinks = true,
	allcolors = blue
}

\usepackage{times}

\makeatother

\usepackage{babel}

\begin{document}

\title{Real eigenvalues are determined by the recursion of eigenstates}
\author{Tong Liu}
\affiliation{School of Science, Nanjing University of Posts and Telecommunications, Nanjing 210003, China}%
\author{Youguo Wang}
\thanks{wangyg@njupt.edu.cn}
\affiliation{School of Science, Nanjing University of Posts and Telecommunications, Nanjing 210003, China}%
\date{\today}

\begin{abstract}
Quantum physics is generally concerned with real eigenvalues due to the unitarity of time evolution. With the introduction of $\mathcal{PT}$ symmetry, a widely accepted consensus is that, even if the Hamiltonian of the system is not Hermitian, the eigenvalues can still be pure real under specific symmetry.
Hence, great enthusiasm has been devoted to exploring the eigenvalue problem of non-Hermitian systems.
In this work, from a distinct perspective, we demonstrate that real eigenvalues can also emerge under the appropriate recursive condition of eigenstates.
Consequently, our findings provide another path to extract the real energy spectrum of non-Hermitian systems, which guarantees the conservation of probability and stimulates future experimental observations.
\end{abstract}

\maketitle

\section{Introduction}\label{n1}
In the textbook of quantum mechanics, the operator of the observable is supposed to the Hermitian operator, to ensure that eigenvalues are totally real~\cite{Shankar}. 
However, with the breakthrough of quantum theory and the development of experimental technology~\cite{Lindblad,Photonic1,ultracold1}, there are several models of open quantum systems in which non-Hermitian Hamiltonians with complex eigenvalues make perfect sense, and both the real and imaginary parts of those are needed to reproduce the measured absorption and emission spectra~\cite{PT1,PT2}. In fact, it is nowadays well understood that having real eigenvalues is a property of the Hamiltonian related to conservation of the total probability~\cite{PT3}, rather than physical observability.

However, the lifetime of the particle in non-Hermitian systems is considered short due to the imaginary part of the eigenvalue during the dynamical evolution. Therefore, non-Hermitian systems with pure real energy spectra are particularly valuable. The emergence of parity-time ($\mathcal{PT}$) symmetry class~\cite{Bender1,Bender2} provides such a paradigm, and formulates an alternative theory of quantum mechanics in which the mathematical axiom of Hermiticity is replaced by the physically transparent condition.
If the Hamiltonian has the unbroken $\mathcal{PT}$ symmetry, fascinatingly, the eigenvalues are pure real.
Hence the $\mathcal{PT}$ symmetry class describes a class of non-Hermitian systems having conservation of the total probability and unitary time evolution.

A question arises naturally: is there other physical mechanism for generating the real energy spectrum~\cite{Zhang1,Lee1} of non-Hermitian systems? Massive research enthusiasm has being devoted to unveiling the new non-Hermitian class~\cite{Yuce1,Yuce2,Yao1,Yao2}. In this work, we attempt to revealing a new mechanism for real energy spectra of non-Hermitian systems, the core idea is that the recursion of eigenstates of the Hamiltonian can constraint the eigenvalues to be the real or complex numbers. In fact, there are many paradigms in quantum mechanics textbooks demonstrating the properties of eigenstates have indeed great influence on eigenvalues.
For example, in the eigenvalue problem of the quantum harmonic oscillator, it can be proved that the Hermite equation has a polynomial solution (Hermite polynomial), namely the wave function is represented as 
$
\phi(x)=\sum_{n=0}^{\infty} a_n x^n.
$
And the recurrence relation of the coefficient $a_n$ satisfies
$
\frac{a_{n+2}}{a_n}= \frac{2 n+1-E}{(n+2)(n+1)}.
$
The problem is that the solution of eigenstates $\phi(x)$ will inevitably lead to divergence in large $x $ limit, so the truncation must occur for the recursive relation of $a_n$.
The most straightforward way is to set the numerator to zero, namely $2 n+1-E=0$. This leads to
$
E=2 n+1,
$
which means eigenvalues are taken as discrete values rather than continuous values in classical physics.

Analogy to the quantum harmonic oscillator model, the discretization of eigenvalues is constrained by the recursion of eigenstates, it is reasonably deduced that there exists a class of models in which the eigenvalues to be the real or complex numbers can be determined by the recursive relation of eigenstates.
The rest of the paper is organized as follows. In Sec.~\ref{n2}, we theoretically demonstrate in detail how the properties of eigenstates determine the real or complex eigenvalues through a simple model. In Sec.~\ref{n3}, we validate the theoretical results through numerical simulations. In Sec.~\ref{n4}, we provide some prospects for more models, and point out that eigenvalues determined by the recursion of eigenstates hold in general. 
In Sec.~\ref{n5}, we make a summary of the paper.

\section{Model and real eigenvalues}\label{n2}
Let's first consider a simple model. Simplicity means that the solution does not require complicated mathematical skills, more relevantly, it can be regarded as a paradigm to grasp the physical picture intuitively. Previous efforts~\cite{Jazaeri,Longhi1} have been made to numerically and semi-analytically solve this model, whereas we attempt to obtain the eigenvalues in an analytical sense. The difference Schr\"{o}dinger equation of the system can be written as
\begin{equation}\label{eq1}
\psi_{n+1}+\psi_{n-1}+V \exp[i (-2\pi\alpha n + \theta)] \psi_n = E \psi_n,
\end{equation}
with the periodic boundary conditions
\begin{equation}
\psi_{n+L}= \psi_n,
\end{equation}
where $V$ is the complex potential strength, $E$ is the eigenvalue of systems, and $\psi_n$ is the amplitude of wave function at the $n$-th lattice. 
We choose to unitize the nearest-neighbor hopping amplitude and a typical choice for irrational parameter is $\alpha=(\sqrt{5}-1)/2$, $\theta$ is the phase factor and generally not zero. Obviously, the complex potential $\exp[i (-2\pi\alpha n + \theta)]$ doesn't satisfy $V(n)=V^*(-n)$ in the discrete lattice, hence the model is independent of $\mathcal{PT}$ symmetry.
This model describes a ring with $L$ sites, where the system size $L$ should be chosen extremely large so that
the irrational number $\alpha$ can be approximated as a rational $\alpha \simeq p/q$ with $p$, $q$ being irreducible integers.
Then we can utilize the discrete Fourier transform
\begin{equation}
\phi_k= \frac{1}{\sqrt{L}} \sum_{n=1}^L \psi_n \exp(2 \pi i \alpha  n k),
\end{equation}
the eigenvalue equation of Eq.~(\ref{eq1}) can be transformed into the momentum space,
\begin{equation}\label{eq2}
\exp(i \theta) \phi_{k-1} + \frac{2}{V} \cos (2 \pi \alpha k) \phi_k = \frac{E}{V} \phi_k,
\end{equation}
there is a multiplying factor $\exp(i \theta)$ in the term $\phi_{k-1}$, however this factor is readily suppressed under the
gauge transformation $\phi_k \rightarrow \exp(i k \theta) \phi_k$.
 
According to Eq.~(\ref{eq2}), an initial wave function solution can be written as
\begin{equation}
\frac{\phi_k}{\phi_{k-1}} \propto
\left\{
\begin{array}{cc}
0 & k <0 \\
1  & k=0 \\
\frac{V}{E-2 \cos (2 \pi \alpha k)} & k > 0.
\end{array}
\right.
\end{equation}
Then the recursive relation of wave function can be written as
\begin{equation}\label{eq3}
\left| \frac{\phi_{k}}{\phi_{0}} \right|=\prod_{k=1}^{L} \left|\frac{V}{E-2 \cos (2 \pi \alpha k)}\right|.
\end{equation}
Obviously, if we know the limit of Eq.~(\ref{eq3}), we can obtain the eigenvalue $E$. 
Let the Ansatz of the normalized wave function be $\left|\phi_{k}\right| \equiv \left|\phi_{0}\right|\exp(- \gamma_m k)$ with $\gamma_m \geq 0$. When $\gamma_m<0$, $\left|\phi_{k}\right|$ cannot be normalized. Then we equivalently transform the Ansatz as
\begin{equation}\label{eq3r}
\gamma_m(E)=- \lim_{k \rightarrow \infty} \frac{1}{k} \ln \left| \frac{\phi_{k}}{\phi_{0}} \right|.
\end{equation}
Substitute Eq.~(\ref{eq3}) into Eq.~(\ref{eq3r}), by utilizing the Weyl's equidistribution theorem~\cite{Longhi2}, the summation can be transformed into the integral,
\begin{equation}\label{eq4}
\begin{aligned}
\gamma_m(E)
&= \lim_{L \rightarrow \infty} \frac{1}{L} \sum_{k=1}^{L} \ln \left| \frac{E-2 \cos(2 \pi \alpha k)}{V} \right|\\
&= \ln \left( \frac{1}{V} \right) + \frac{1}{2 \pi}  \int_{0}^{2 \pi} \ln \left| E - 2 \cos(\tilde{k})\right| d\tilde{k}.\\
\end{aligned}
\end{equation}
Obviously, to obtain the eigenvalue $E$, we need to know the value of $\gamma_m$. In fact, $\gamma_m$ has the explicit physical significance, which is Lyapunov exponent in momentum space.

To obtain $\gamma_m$, we can utilize a famous formula, which has been obtained initially for random systems by Thouless and can be used without any change for non-random systems. Namely, Lyapunov exponent can be related to the density of states by
\begin{equation}\label{eq5}
\gamma(E) =\int d E^{'} \ln| E-E^{'}|\rho(E^{'}),
\end{equation}
which is dubbed Thouless formula~\cite{Thouless}. For non-Hermitian tight-binding lattices with nearest-neighbor hopping, provided that the hopping amplitudes are symmetric, a similar relation can be established~\cite{Longhi1}. The advantage of this formula is able to connect Lyapunov exponent in position space $\gamma$ and Lyapunov exponent in momentum space $\gamma_m$.
The definition of density of states is the number of quantum states with energy ranging from $E$ to $E+\Delta E$. Under Fourier transform, the eigenvalues of Eq.~(\ref{eq1}) and Eq.~(\ref{eq2}) remain unchanged. Hence Eq.~(\ref{eq1}) and Eq.~(\ref{eq2}) have the same density of state,
\begin{equation}\label{dos}
\rho(E)=\rho_m(\frac{E}{V}),
\end{equation}
then substitute Eq.~(\ref{dos}) into Eq.~(\ref{eq5}),
\begin{equation}\label{eq6}
\begin{aligned}
\gamma(E) &=\int d E^{'} \ln| E-E^{'}|\rho_m(\frac{E^{'}}{V}),\\
&=\int d E^{'} \ln| \frac{E-E^{'}}{V}|\rho_m(\frac{E^{'}}{V})+\ln(V),\\
&=\gamma_m(\frac{E}{V})+\ln(V).
\end{aligned}
\end{equation}
From Eq.~(\ref{eq6}), if we obtain $\gamma$, $\gamma_m$ can also be naturally obtained.

As regard to $\gamma$, we refer to the transfer matrix method, which can be used for the analysis of the wave propagation in classical or quantum systems. The growth or decay of the propagation is governed by the Lyapunov spectrum of the product of transfer matrices. For the one-dimensional nearest-neighbor hopping model, the transfer matrix is two-dimensional, the nonzero or zero values of Lyapunov exponents are utilized to measure whether waves are localized at a certain location or spread throughout the entire space. The specific transfer matrix of Eq.~(\ref{eq1}) can be written as
\begin{equation}
T(\theta)=\left(
\begin{array}{cc}
 E -  V \exp[i (2\pi\alpha + \theta)]&  -1 \\
1 & 0 \\
\end{array}
\right),
\end{equation}
according to Avila's theory~\cite{Avila}, let us complexify the phase $\theta \rightarrow \theta -i \vartheta$, and let $\vartheta \rightarrow +\infty$, direct computation shows that
\begin{equation}
T(\theta -i \vartheta)=\exp(\vartheta)\exp[i(2\pi\alpha+ \theta)]\left(
\begin{array}{cc}
 -V &  0 \\
0 & 0 \\
\end{array}
\right) + o(1).
\end{equation}
Thus we have  $\gamma \left( E,\vartheta\right)=\lim\limits_{n\to \infty}\frac{1}{ n}\ln  \|T_n(\theta -i \vartheta)\|= \vartheta +\ln(V) +o(1)$. Note
 $\gamma (E,\vartheta)$ is a convex, piecewise linear function of $\vartheta$ with their
slopes being integers, if the energy $E$ belongs to the spectrum, then
\begin{equation}
\gamma \left( E,\vartheta\right)= \max\{0,\ln(V)+\vartheta\}, ~~\forall \vartheta \geq 0,
\end{equation}
consequently, we have $\gamma \left( E\right)= \max\{0,\ln(V)\}$ by setting $\vartheta=0$.

According to Eq.~(\ref{eq6}), also note that $\gamma$ and $\gamma_m$ are independent of $E$, we obtain the Lyapunov exponent in momentum space 
\begin{equation}
\gamma_m(E)= \max\{0,\ln \left( \frac{1}{V} \right) \}.
\end{equation}
It is obviously, when $V<1$, the real space Hamiltonian is in the extended phase, whereas the momentum space Hamiltonian is in the localized phase; while $V>1$, the real space Hamiltonian is in the localized phase, whereas the momentum space Hamiltonian is in the extended phase.
It should be emphasized that Lyapunov exponent cannot directly give the information of eigenvalues, taking any value of $E$ satisfies the formula $\gamma=\ln(V)$, however, the eigenvalue of the system is certainly not arbitrary.

Consequently, when $ \gamma_m(E)= \ln \left( \frac{1}{V} \right) $, namely $0<V\leq1$, from Eq.~(\ref{eq4}) it can be obtained $\frac{1}{2 \pi}  \int_{0}^{2 \pi} \ln \left| E - 2 \cos(\tilde{k})\right| d\tilde{k}=0.$ From the above integral equation, we can obtain that the set of all allowed eigenvalues is 
\begin{equation}
E = 2 \cos(\tilde{k})
\end{equation}
with $0 \leq \tilde{k}< 2 \pi.$
This result has important physical consequences, it means that when $0<V\leq1$, all eigenvalues are real numbers.

When $ \gamma_m(E)= 0 $, namely $V>1$, from Eq.~(\ref{eq4}) it can be obtained
$\frac{1}{2 \pi}  \int_{0}^{2 \pi} \ln \left| E - 2 \cos(\tilde{k})\right| d\tilde{k} =\ln \left(V\right).$
This is Dini Integral, which is a logarithmic integral with
important applications in mathematical physics and engineering, one first evaluated in 1878 by the Italian mathematician Ulisse Dini.
According to the conclusions of Dini Integral, we can obtain that the set of all allowed eigenvalues is 
\begin{equation}
E = V e^{i \tilde{k}}+\frac{e^{-i \tilde{k}}}{V}
\end{equation}
with $0 \leq \tilde{k} < 2 \pi.$
At the phase transition point $V=1$, two sets of eigenvalues can be connected smoothly. In the above derivation, we focus on the integral equation of the Lyapunov exponent in the momentum space, essentially the recursion of eigenstate in the momentum space, to determine the value range of eigenvalues.

\section{Numerical verification}\label{n3}
\begin{figure}
  \centering
  \includegraphics[width=0.5\textwidth]{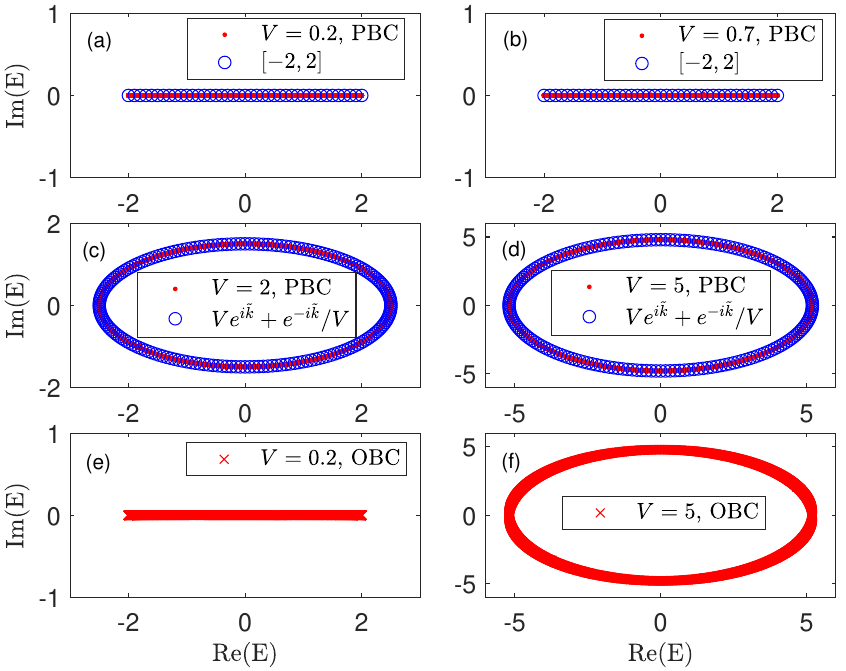}\\
  \caption{(Color online) The eigenvalues of Eq.~(\ref{eq1}) are illustrated, the abscissa is the real part of the eigenvalues, and the ordinate is the imaginary part. The total number of sites is set to be $L=2000$.  The red dots represent numerical solutions under periodic boundary conditions, the blue circles represent theoretical values, and the red ``x'' represent numerical solutions under open boundary conditions. As shown in (a) and (b), when $V<1$, the system host the pure real energy spectrum $[-2, 2]$. While $V>1$ [(c) and (d)], the energy spectrum of the system forms a closed loop, and the numerical solutions are in good agreement with the theoretical predicted values. (e) and (f) show that the spectra are not affected by the boundary conditions.}
  \label{fig1}
\end{figure}

\begin{figure}
  \centering
  \includegraphics[width=0.5\textwidth]{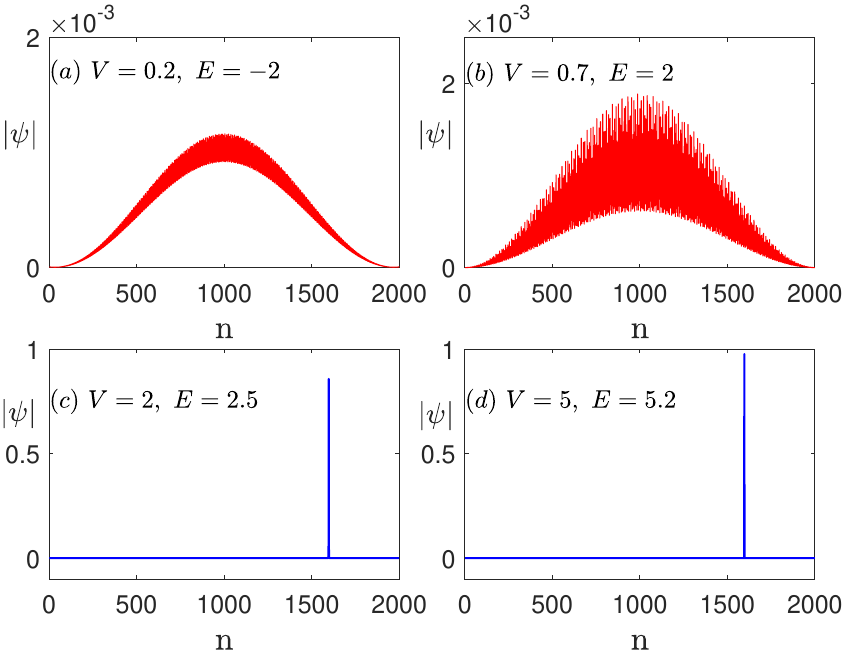}\\
  \caption{(Color online)  The absolute value of eigenstates of Eq.~(\ref{eq1}) under periodic boundary conditions for typical eigenvalues. The total number of sites is set to be $L=2000$. (a) and (b) demonstrate that the general real eigenvalues correspond to extended states when $V<1$; (c) and (d) demonstrate that some special real eigenvalues correspond to localized states when $V>1$.}
  \label{fig2}
\end{figure}

To support the above analytical result, we now present the numerical verification, namely directly diagonalize Eq.~(\ref{eq1}) to obtain the eigenvalues and eigenstates. In Fig.~\ref{fig1}(a) and (b), numerical results under periodic boundary conditions demonstrate that when the potential strength $V<1$, all eigenvalues of the system are filled with intervals $[-2, 2]$, therefore, they are pure real. Thus, the conserved evolution probability of the system is guaranteed, just as the $\mathcal{PT}$ unbroken system. While $V>1$, as shown in Fig.~\ref{fig1}(c) and (d), the imaginary part of eigenvalues is no longer limited to 0, whereas the real and imaginary parts form a closed loop, satisfing the expression $E = V e^{i \tilde{k}}+\frac{e^{-i \tilde{k}}}{V}$ with $0 \leq \tilde{k} < 2 \pi.$ 

Considering non-Hermitian skin effect~\cite{Yao1,Yao2}, real energy spectra can be induced by open boundary conditions, we further perform the numerical simulation for Eq.~(\ref{eq1}) under open boundary conditions. As shown in Fig.~\ref{fig1}(e) and (f), and as compared to Fig.~\ref{fig1}(a) and (d), the energy spectra remain unchanged under both two boundary conditions, which demonstrates that the eigenvalue problem of Eq.~(\ref{eq1}) is independent of non-Hermitian skin effect.
Thus, all numerical results are completely consistent with the theoretical predictions, which confirm the validation of our theory.

Resulting from that all eigenstates of the system are extended for $V<1$, there exists a major misunderstanding that extended states ($V<1$) correspond to real eigenvalues, and localized states ($V>1$) correspond to complex eigenvalues, as shown in Fig.~\ref{fig2}.
Here we should clarify that some special localized states can also host real eigenvalues. When $V>1$, with regard to the $\tilde{k}=0$ eigenstate, the eigenvalue is $E_0 =V + \frac{1}{V}>2$, which means that the localized eigenstates with eigenvalues $V + \frac{1}{V}$ are also real numbers, as shown in Fig.~\ref{fig2}(c) and (d). Hence, the corresponding principle mentioned above is no longer valid, which is consistent with that the real-complex spectrum transition does not originate from some symmetry breaking. The suitable recursive relation can lead to the appearance of the real energy, whether the eigenstate is an extended state or a localized state.

\section{prospect in more modles}\label{n4}
In fact, real eigenvalues in many complicated models~\cite{liu1,liu2,liu3,xia} also originated from the recursion of eigenstates. However, due to the complexity of models, obtaining rigorous mathematical solutions is very difficult. To illustrate the generality of the framework, we briefly introduce two models, and discuss some semi-analytical solutions.

Firstly, we introduce the following difference equation~\cite{liu3}, which is written as
\begin{equation}\label{critical}
\psi_{n+1}+\psi_{n-1}+V i \tan(2\pi\alpha n+\theta) \psi_n = E \psi_n,
\end{equation}
where the meaning of each parameter is the same as Eq.~(\ref{eq1}), except that the potential is replaced by $i \tan(2\pi\alpha n+\theta)$ from $\exp[i (-2\pi\alpha n+\theta)]$.
Through Fourier transformation, the dual equation of Eq.~(\ref{critical}) in momentum space is written as
\begin{equation}\label{dualcritical}
{\phi_{k+1}} =\frac{-2\cos [2 \pi (k-1) \alpha]+V +E}{2\cos [2 \pi (k+1)\alpha]+V -E } \phi_{k-1}.
\end{equation}
From Eq.~(\ref{dualcritical}), Lyapunov exponent $\gamma_m$ can be obtained,
\begin{equation}\label{dualcritical1}
\gamma_m(E)=\frac{1}{2 \pi} \int_0^{2 \pi}\ln g^{(1)}-\ln g^{(2)} d \theta,
\end{equation}
where $g^{(1)}=|-2\cos (2 \pi  \theta)+V +E |,g^{(2)}=|2\cos (2 \pi  \theta)+V -E |$.
Then we get the explicit expression for Lyapunov exponent in positon space $\gamma$ by utilizing Avila's global theory~\cite{Avila},
\begin{equation}
\begin{aligned}
\gamma(E) = \max\{&\operatorname{arcosh} \frac{\left| E+V+2\right|+\left| E+V-2\right|}{4},\\
&\operatorname{arcosh} \frac{\left|E-V+2\right|+\left|E-V-2\right|}{4}\} .
\end{aligned}
\end{equation}
Unfortunately, due to the complexity of this model, we are unable to obtain the explicit expression of $\gamma_m$ through Thouless formula. Alternatively, we make $\gamma(E)=0$, and obtain that $E$ is within the region $ [V-2, 2-V]$; we make $\gamma(E)>0$, and obtain that $E$ is within the region $\{i y~|~y\in \mathbb{R}^{*}\}$ ($V\leq 2$) or $\{i y~|~y\in \mathbb{R}\}$ ($V > 2$). Then, we substitute these guessed energies ``E'' into Eq.~(\ref{dualcritical1}), and find that when $E\in[V-2, 2-V]$, $\gamma_m(E)>0$, while $E$ is a pure imaginary number, $\gamma_m(E)=0$, all detailed calculations can be found in Ref.~\cite{liu3}. Since the energy ``E'' exactly satisfies the duality relation ($\gamma=0, \gamma_m>0$ and $\gamma>0, \gamma_m=0$) indicated by Thouless formula between Lyapunov exponent in position space and momentum space, we conjecture that ``E'' is the eigenvalue of Eq.~(\ref{critical}).

In addition to quasiperiodic models, real eigenvalues determined by the recursion of eigenstates is also applicable to random disordered systems. 
A paradigm of non-Hermitian random disorder is Hatano-Nelson model~\cite{HN}, which originated from the study of the pinning of flux lines by random columnar defects in a superconductor. In the clean limit (no random impurities), the model is well known as the non-Hermitian skin effect due to the imaginary gauge field $h$, the eigenvalues form an ellipse on the complex plane under periodic boundary conditions, and the corresponding eigenstates are extended.
With increase of concentration of random impurities, real eigenvalues emerge at the edge of spectra and the corresponding eigenstates are localized, 
while complex eigenvalues at the center of spectra also correspond to extended eigenstates. On the surface, it seems that Hatano-Nelson model violates the principle of real eigenvalues corresponding to extended eigenstates for quasiperiodic models. Actually, numerous works in physics community~\cite{HN1,HN2,HN3} have shown that the emergence of real energy spectra still stems from Lyapunov exponent of the eigenstate for Hatano-Nelson model. Unfortunately, a complete and rigorous calculating result is still missing. Nevertheless, some mathematical references~\cite{HN4,HN5} demonstrate  that the behaviour of the eigenvalues depends crucially on the Lyapunov exponent associated to the Hermitian operator. The mathematical skills of relational papers are quite advanced, and we directly quote their conclusions. Making the potential of random impurities has the uniform distribution $[-1,1]$, there exist two critical values $0<h_1<h_2$ and the following hold: (i) when $0\leq h<h_1$, the eigenvalues of Hatano-Nelson model are totally real;
(ii) when $h_1< h<h_2$, some of the eigenvalues remain real, while others form a smooth curve on the complex plane; (iii) when $h_2< h$, all eigenvalues become complex. An intuitive understanding is that when the imaginary gauge field $h=0$, the system is the Hermitian Anderson model, and the eigenvalues are pure real accompanied by localized eigenstates. While $h$ gradually increases, the eigenvalues corresponding to the localized state remain real, nevertheless, the complex eigenvalues induced by non-Hermitian effect correspond to the delocalization of the eigenstates. Thus, the recursions of eigenstates are still closely linked with eigenvalues.

It is worth mentioning that, the exact solution of eigenvalues for non-Hermitian disordered and quasi-disordered systems is a huge challenge, the complete clarification of these problems depend on the progress of future mathematical tools.

\section{Summary}\label{n5}
In summary, we provide, for the first time analytically, an example of pure real energy spectrum originated from the recursive relation of eigenstates, which is different from the known physical mechanism, such as the $\mathcal{PT}$ symmetry. As long as the recursion of the eigenstate is determined, the eigenvalue of the system may have a pure real energy spectrum, which means that the system can undergo unitary time evolution. In addition, we need to emphasize that extended states of the system lead to the real eigenvalues in most cases, however, this is not a necessary condition for the eigenvalue to be a real number, and localized states can also produce the real eigenvalues. Finally, we provide the prospect that eigenvalues determined by the recursion of eigenstates are widely present in various systems. Our discovery of this non-Hermitian phenomenon promotes the realm of the eigenvalue problem in non-Hermitian quantum theory towards a new avenue, and these findings are expected to be of great interest to the broad community.

\begin{acknowledgments}
This work was supported by  the National Natural Science Foundation of China (Grant No. 62071248), the Natural Science Foundation of Jiangsu Province (Grant No. BK20200737), NUPTSF (Grants No. NY220090 and No. NY220208), the Innovation Research Project of Jiangsu Province (Grant No. JSSCBS20210521), and China Postdoctoral Science Foundation (Grant No. 2022M721693).
\end{acknowledgments}


\end{document}